\documentclass[openacc]{rstransa}%
\usepackage{upgreek}
\usepackage{siunitx}

\newcommand{\solar}{\odot}
\titlehead{Review}

\begin{document}

\title{Searches for beyond-standard-model physics with astroparticle physics instruments}

\author{%
M. Ackermann$^{1}$ and K. Helbing$^{2}$}

\address{$^{1}$Deutsches Elektronen-Synchrotron DESY, Platanenallee 6, 15738 Zeuthen, Germany\\
$^{2}$Dept. of Physics, University of Wuppertal, 42119 Wuppertal, Germany
}
\subject{Astroparticle physics, Neutrino properties, Fundamental particles and interactions}

\keywords{Neutrinos, Gamma rays, Beyond-standard-model physics}

\corres{Markus Ackermann\\
\email{markus.ackermann@desy.de}\\
Klaus Helbing\\
\email{helbing@uni-wuppertal.de}}

\begin{abstract}

Many instruments for astroparticle physics are primarily geared towards multi-messenger astrophysics, to study the origin of cosmic rays (CR) and to understand high-energy astrophysical processes. Since these instruments observe the Universe at extreme energies and in kinematic ranges not accessible at accelerators these experiments provide also unique and complementary opportunities to search for particles and physics beyond the standard model of particle physics. In particular, the reach of IceCube, Fermi and KATRIN to search for and constrain Dark Matter, Axions, heavy Big Bang relics, sterile neutrinos and Lorentz Invariance Violation (LIV) will be discussed. The contents of this article are based on material presented at the Humboldt-Kolleg "Clues to a mysterious Universe - exploring the interface of particle, gravity and quantum physics" in June 2022.
\end{abstract}

\begin{fmtext}
\end{fmtext}

\maketitle

\section{Introduction} %
Neutrino oscillations and dark matter are clear indications that the standard model of particle physics is incomplete. Despite this evidence, searches at accelerators for new particle types and new interactions have not shown results so far up to TeV masses. Astroparticle physics experiments\footnote{We use a broad definition of astroparticle physics in this manuscript including any particle physics instrumentation that does not require an accelerator.} go beyond the reach of accelerator experiments to higher and lower energies.
In this paper we describe the current status of KATRIN, IceCube and Fermi, explore the prospects of these instruments for measuring physics observables connected to beyond-Standard-Model (BSM) hypotheses, and discuss research-and-development projects that may further extend their reach. It is structured as follows: \autoref{sec:instruments_observations} introduces the instruments, their key observations and how they enable constraints on or discoveries of BSM physics; \autoref{sec:new_physics} summarizes a selection of specific BSM physics constraints obtained by aforementioned instruments;  \autoref{sec:future} provides a glimpse into the future with an overview of related planned astroparticle physics instruments.

\section{Instruments and key observations}
\label{sec:instruments_observations}
\subsection{IceCube} %
The IceCube neutrino observatory~\cite{IceCube:2016zyt}, located near Amundsen-Scott base at the geographic South Pole in Antarctica, is the world's most sensitive neutrino telescope to date. 
About one cubic kilometer of glacial ice is instrumented with 5160 optical sensors that detect the faint Cherenkov light of charged particles traversing the ice, such as muons produced in charged-current (CC) interactions of muon neutrinos, or particle showers originating from deep inelastic scattering of neutrinos of all flavors. 
The optical sensors, deployed at depths between 1450\,m and 2450\,m below the surface along 86 cables are interspersed throughout the volume with a typical distance of about 125\,m between cables, and 17\,m between adjacent sensors on a single cable. IceCube also contains a surface detector (IceTop) equipped with identical optical sensors that cover the footprint of the in-ice detector, allow CR composition studies, and aid in the separation of neutrinos from CR air showers. Constructed during the years from 2005 and 2011, IceCube has been fully operational and is continuously taking data since then with an uptime in excess of 99\%.  

Arrival direction, energy, and flavor reconstruction for the incident neutrinos is performed based on the timing and amplitude information collected by the optical sensors, called digital optical modules (DOMs). The same information is also used to distinguish neutrino interactions from atmospheric backgrounds induced by cosmic-ray air showers. 
Typical particle signatures, distinguished in IceCube analysis, are: a) \emph{tracks}, elongated light patterns from muons traversing the detector volume; b) \emph{cascades}, hadronic and electromagnetic particle showers contained in a small volume inside the detector; c) \emph{starting tracks}, the combination of a track and a cascade due to a CC neutrino interaction inside the instrumented volumel d) \emph{double cascades}, from high-energy (E$\,\gtrsim$\,100\,TeV) CC tau neutrino interactions and the subsequent decay of the tau lepton. 
Some hypothesized beyond-standard-model particles would be identified in IceCube by their non-standard signatures that are distinct from neutrinos or atmospheric backgrounds, e.g., by a velocity significantly slower than $c_0$, the vacuum light speed, by light yields that point to fractional charges, or by a double track signature. 

Each particle signature has its unique values and limitations. The arrival direction of tracks can be reconstructed to sub-degree accuracy~\cite{IceCube:2016zyt}, while the angular resolution is of O(10$^\circ$) for cascades. In contrast, the energy resolution of cascades detected in IceCube is $\sim$15\%~\cite{IceCube:2013dkx}, while the measurement of the energy loss of tracks inside the instrumented volume only yields a lower limit on the muon and neutrino energies. Double cascades are specific to tau neutrino interactions and play an important role for constraining the flavor composition of the astrophysical neutrino flux. The energy threshold for the bulk of the detector array corresponds to about 100~GeV of deposited energy, i.e., the cumulative energy loss of the neutrino-induced muon or particle shower inside the instrument. A denser configuration of optical sensors in the center of IceCube (named \emph{DeepCore}) yields a lower detection threshold of about 10~GeV for neutrino interactions in this volume, enabling the study of neutrino oscillation parameters with atmospheric neutrinos in IceCube, but also significantly increasing the sensitivity to neutrino signals from dark matter annihilation for WIMP masses below few hundreds of GeV~\cite{IceCube:2011ucd}.      

\subsection{Fermi Gamma-ray Space Telescope} %

The \emph{Fermi} Gamma-ray Space Telescope, launched in 2008 into a low Earth orbit, surveys the gamma-ray sky with unprecedented sensitivity for the last 15 years. The satellite contains two instruments. The main payload is the Large Area Telescope (LAT), surveying the entire gamma-ray sky at energies from about 50\,MeV to 1\,TeV~\cite{Fermi-LAT:2009ihh}. A particle tracking detector, consisting of multiple layers of Si-strip detectors, interspersed with Tungsten foils records the trajectories of electron-positron pairs produced from gamma-rays interacting in the detector. Their energy is measured in a calorimeter system consisting of CsI scintillator crystals mounted below the tracking detector. Both detectors are shielded against incoming charged cosmic rays by an anti-coincidence veto system made of plastic scintillator tiles that surround the tracker and calorimeter on five sides.

The resolution for measuring the energy of the incoming photons is better than 10\% over a wide range in energy. The angular resolution depends strongly on the photon energy and ranges from few degrees at 100\,MeV to O(0.1$^\circ$) at 10\,GeV and above. The LAT has a wide field-of-view and observes at any instant a solid angle of 2.4\,sr, allowing the entire sky to be observed every three hours\footnote{between 2008 and 2018, modified survey pattern afterwards. See~\url{https://fermi.gsfc.nasa.gov/ssc/observations/types/post_anomaly/} for more information.} An important aspect of the LAT mission that is reflected in its design is the search for the specific signatures of gamma rays produced in the annihilation of weakly interacting massive particles (WIMPs) in various regions of the sky. It is also sensitive to certain proposed beyond-standard-model physics effects through the observations of the spectral shapes of astrophysical sources and the light curves of transients.

The second instrument on the Fermi satellite is the Gamma-ray Burst Monitor (GBM) designed to detect and localize short gamma-ray transients in two energy bands (8\,keV -- 1\,MeV, 200\,keV -- 40\,MeV)~\cite{2009ApJ...702..791M}. Its field-of-view is the entire sky that is not occulted by the shadow of the Earth. It routinely detects hundreds of gamma-ray bursts (GRB) per year, extremely energetic transients of seconds to minutes duration associated with particular types of supernova explosions and compact object mergeres (see, e.g., ~\cite{Meszaros:2001vi} for a review). In 2017, the GBM was one of the instruments that observed the electromagnetic counterpart to the binary neutron star merger GW170817A in coincidence with the gravitational wave signal~\cite{LIGOScientific:2017vwq,LIGOScientific:2017ync}, marking the first and only direct confirmation of the connection between compact object mergers and GRBs to date.

\subsection{KATRIN} %

\begin{figure}
    \centering
    \includegraphics[width=\textwidth]{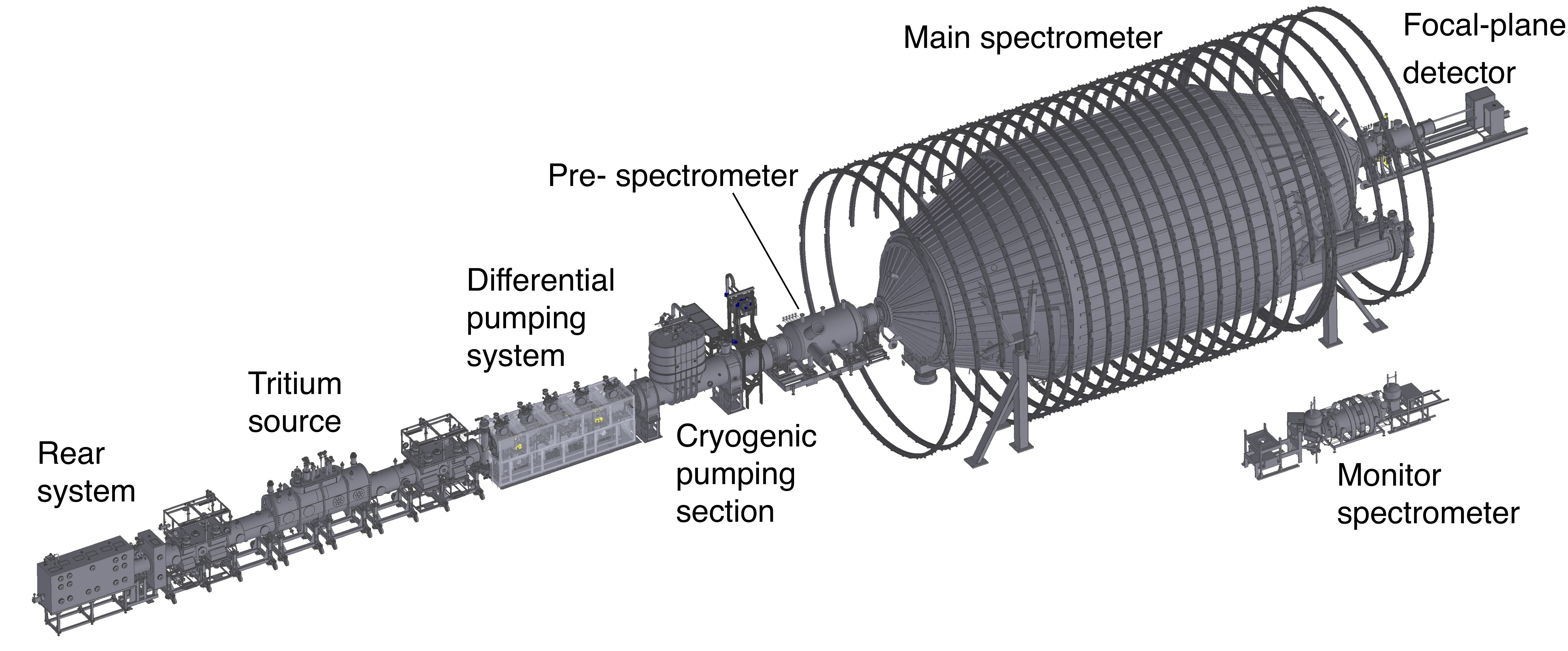}
    \caption{The KATRIN beamline. Figure adapted from~\cite{KATRIN:2022ayy}}
    \label{fig:beamline}
\end{figure}

The KATRIN experiment is designed to measure the mass of the electron antineutrino with sub-eV precision by examining the spectrum of electrons emitted from the beta decay of tritium. 
The experiment is a large scale experiment carried out at the Karlsruhe Institute of Technology in Germany scaling up the size of previous experiments by an order of magnitude including a much more intense gaseous tritium source. 
The primary goal of the experiment is to determine the absolute mass scale of the neutrino in a direct and model-independent measurement.
After a first tritium commissioning campaign in 2018, the experiment is collecting data since 2019, and in its first two measurement campaigns has already achieved a sub-eV sensitivity. After 1000 days of data-taking, KATRIN's design sensitivity is 0.2 eV at the 90\% confidence level.  

$\upbeta$ decays are provided by a high-luminosity, windowless, gaseous, cryogenic, molecular tritium source, while the $\upbeta$ energies are analyzed by a MAC-E-filter spectrometer with \si{\electronvolt}-scale filter width.
The \SI{70}{\metre} KATRIN beamline is designed to perform high-precision energy analysis of $\upbeta$ electrons from a high-luminosity, gaseous T$_2$ source. 
The following is an overview of the apparatus (Fig.~\ref{fig:beamline}). 
Briefly, T$_2$ gas is purified in a tritium loop system, which continuously delivers cold T$_2$ gas to the center of the source system. 
The gas diffuses to both ends of the source-system cryostat, where the first pumping stages are located. 
A fraction of the T$_2$ molecules experience $\upbeta$ decay during their flight within the source beam tube. 
Of the resulting $\upbeta$s, those that are emitted in the downstream direction are guided by magnetic field lines through the chicanes of two consecutive pumping stages that together reduce the flow of neutral tritium by some \num{14}~orders of magnitude. 
Past the pumping systems, the $\upbeta$s reach a pair of tandem spectrometers designed according to the principle of magnetic adiabatic collimation with electrostatic filtering (MAC-E filters). 
A MAC-E filter uses magnetic-field gradients to collimate the $\upbeta$-electron flux, allowing a longitudinal retarding potential $U$ to set a threshold on the total kinetic energy of the $\upbeta$s. 
A MAC-E filter thus acts as an integrating high-pass filter, with a characteristic relative filter width set by the ratio of the minimum to the maximum magnetic fields. 
$\upbeta$ electrons with sufficient energy pass through the main spectrometer and are counted in the focal-plane detector.
Calibration and monitoring systems are located in several positions along the beamline; e.g. the forward beam monitor~\cite{Beglarian:2021ubj} in the cryogenic pumping section uses silicon detectors to sample the $\upbeta$ rate at the edge of the flux tube and across. Recently, 
the KATRIN experiment has placed an upper limit on the neutrino mass of $m_\nu <$ \SI{0.8}{\electronvolt} ~\cite{KATRIN:2021uub}. 
Besides this main scientific goal, KATRIN is also sensitive to several BSM hypotheses as discussed in~\autoref{sec:new_physics}.

\subsection{Astrophysical observations} %
\label{sec:astro_observations}

\begin{figure}
    \centering
    \includegraphics[width=0.9\linewidth]{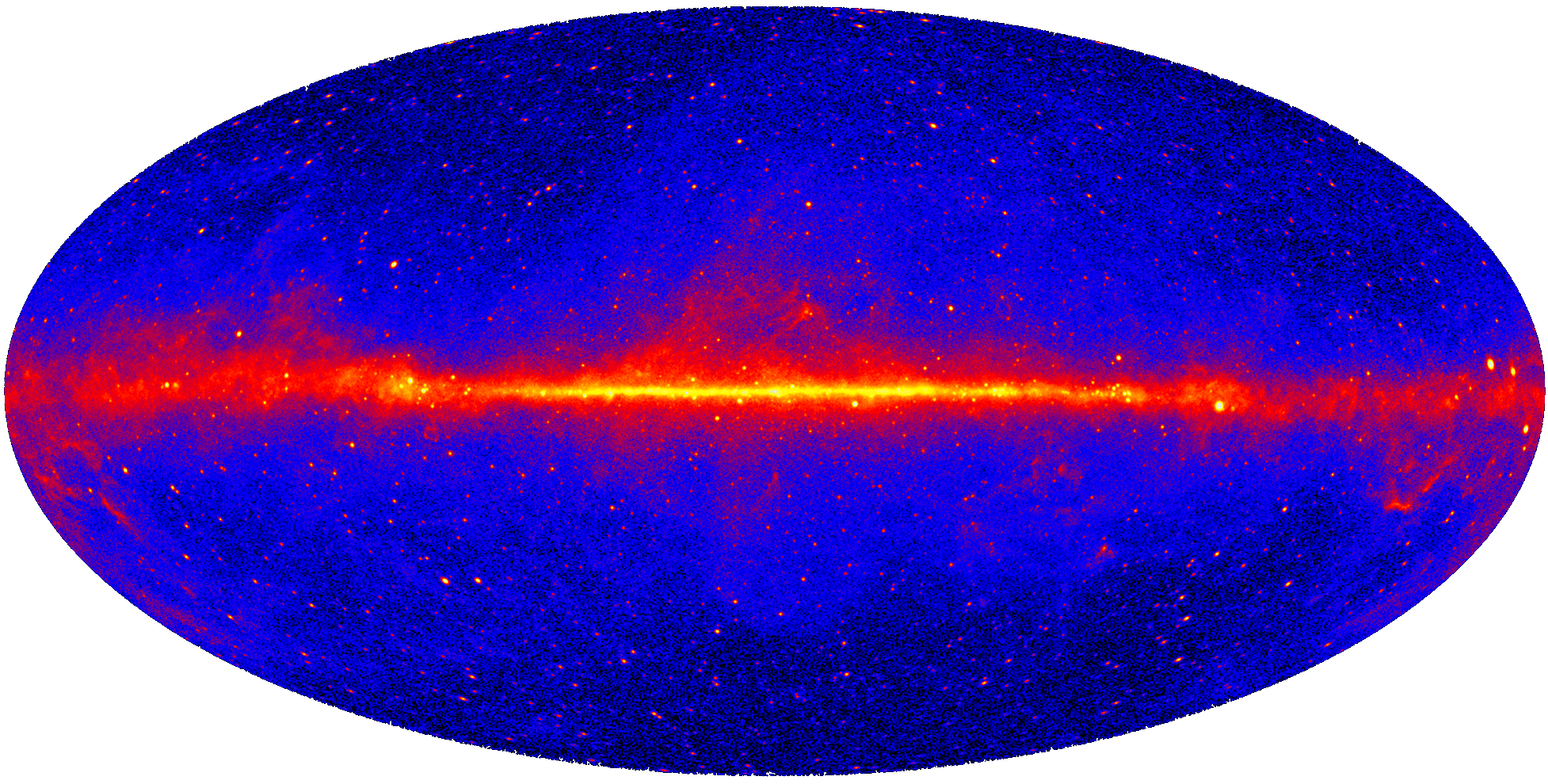}
    \caption{Gamma-ray sky observed above 1~GeV in energy by the Fermi Large Area Telescope. Credit: NASA / DOE / Fermi LAT Collaboration.}
    \label{fig:latsky}
\end{figure}

The Fermi LAT has given us the most detailed picture of sources, transients, and diffuse emission on the gamma-ray sky to date. \autoref{fig:latsky} shows a map of the sky above 1\,GeV in energy corresponding to 12 years of LAT observations. Clearly visible on this skymap are the many sources, both Galactic and extragalactic in origin, as well as the diffuse Galactic emission originating from the interactions of cosmic-ray nuclei and electrons in the interstellar gas and radiation fields~\cite{Fermi-LAT:2012edv}. Also visible are the giant lobes that are commonly associated with an outflow from the central region of the Galaxy, while their detailed production mechanisms are still under debate (see, e.g.,~\cite{Yang:2018arq} for a review). Source catalogs are updated in regular intervals by the Fermi LAT collaboration, the latest such catalog being 4FGL-DR3, based on 12 years of LAT observations~\cite{Fermi-LAT:2022byn}. 6558 gamma-ray sources have been reported in this catalog, of which about two thirds have been identified or associated to known astrophysical sources. Many different populations of Galactic gamma-ray sources contribute. The extragalactic sky, however, is overwhelmingly dominated by blazars -- active galaxies, with their relativistic jet pointing towards Earth. While such systems are rare, they can be observed throughout almost the entire universe up to high redshifts (z$\,\gg\,$2),
due to the already high intrinsic gamma-ray luminosity of the jets being further amplified by Doppler boosting effects. 

In the core of the Milky way an excess of gamma-radiation at few GeV in energy has been observed  that is difficult to reconcile with models of the diffuse emission from cosmic-ray interactions in the interstellar medium. Gamma-ray emission from the annihilation of WIMPs near the center of our Galaxy has been proposed as a potential origin for this radiation, but it remains inconclusive, as several scenarios for an astrophysical origin of this excess remain viable, e.g., emission from unresolved millisecond pulsars in the bulge of the Milky Way (see, e.g.,~\cite{Murgia:2020dzu} for a review).

\begin{figure}
    \centering
    \includegraphics[width=0.7\linewidth]{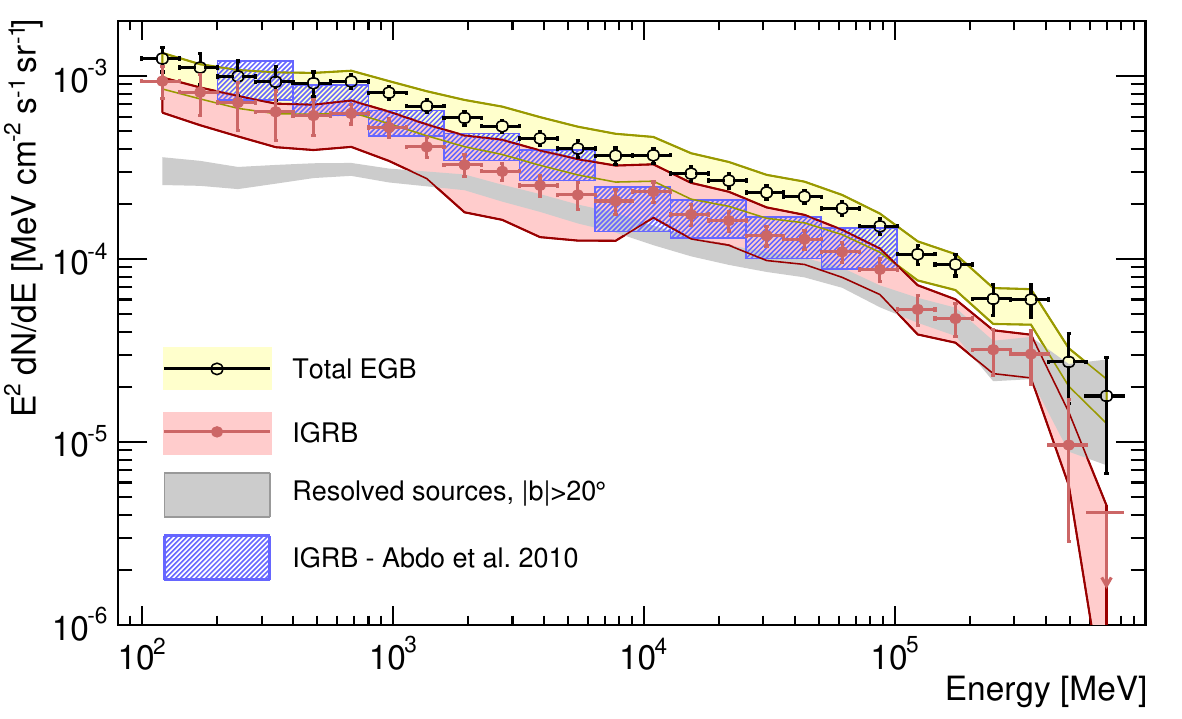}
    \caption{Isotropic (IGRB) and total extragalactic gamma-ray background (EGB) observed by Fermi LAT. The total EGB comprises the sum of the IGRB and the resolved extragalactic sources. An earlier measurement~\cite{Fermi-LAT:2010pat} is shown for comparison. Figure adapted from~\cite{Fermi-LAT:2014ryh}.}
    \label{fig:igrb}
\end{figure}

The extragalactic diffuse gamma-ray emission, also called the isotropic gamma-ray background (IGRB) contains the emission of all unresolved source populations throughout the universe, as the universe is transparent to gamma rays (E$\,\lesssim\,$10\,GeV) up to distances far beyond the peak in star and galaxy formation. \autoref{fig:igrb} shows the latest measurement of the IGRB obtained from Fermi LAT data~\cite{Fermi-LAT:2014ryh}. It is particularly interesting for searches for BSM physics, as any hypothetical process that eventually leads to the production of gamma rays would leave an imprint in the IGRB. Accordingly, the measured IGRB intensity can be used to constrain proposed processes or particle properties.

\begin{figure}
    \centering
    \includegraphics[width=0.9\linewidth]{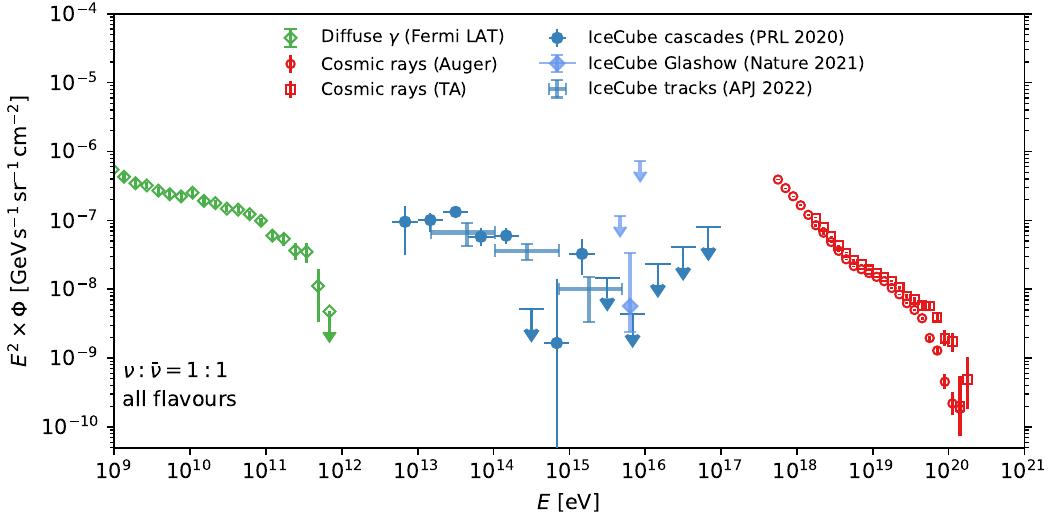}
    \caption{Measurements of high-energy gamma rays by Fermi-LAT~\cite{Fermi-LAT:2014ryh}, high-energy neutrinos by IceCube~\cite{IceCube:2020acn,IceCube:2021uhz,IceCube:2021rpz}, and ultra-high-energy cosmic rays by the Pierre Auger Observatory~\cite{PierreAuger:2021hun} and the Telescope Array~\cite{Ivanov:2020rqn}. Figure adapted from~\cite{Ackermann:2022rqc}.}
    \label{fig:mm_paradigm}
\end{figure}

\begin{figure}
    \centering
    \includegraphics[width=0.6\linewidth]{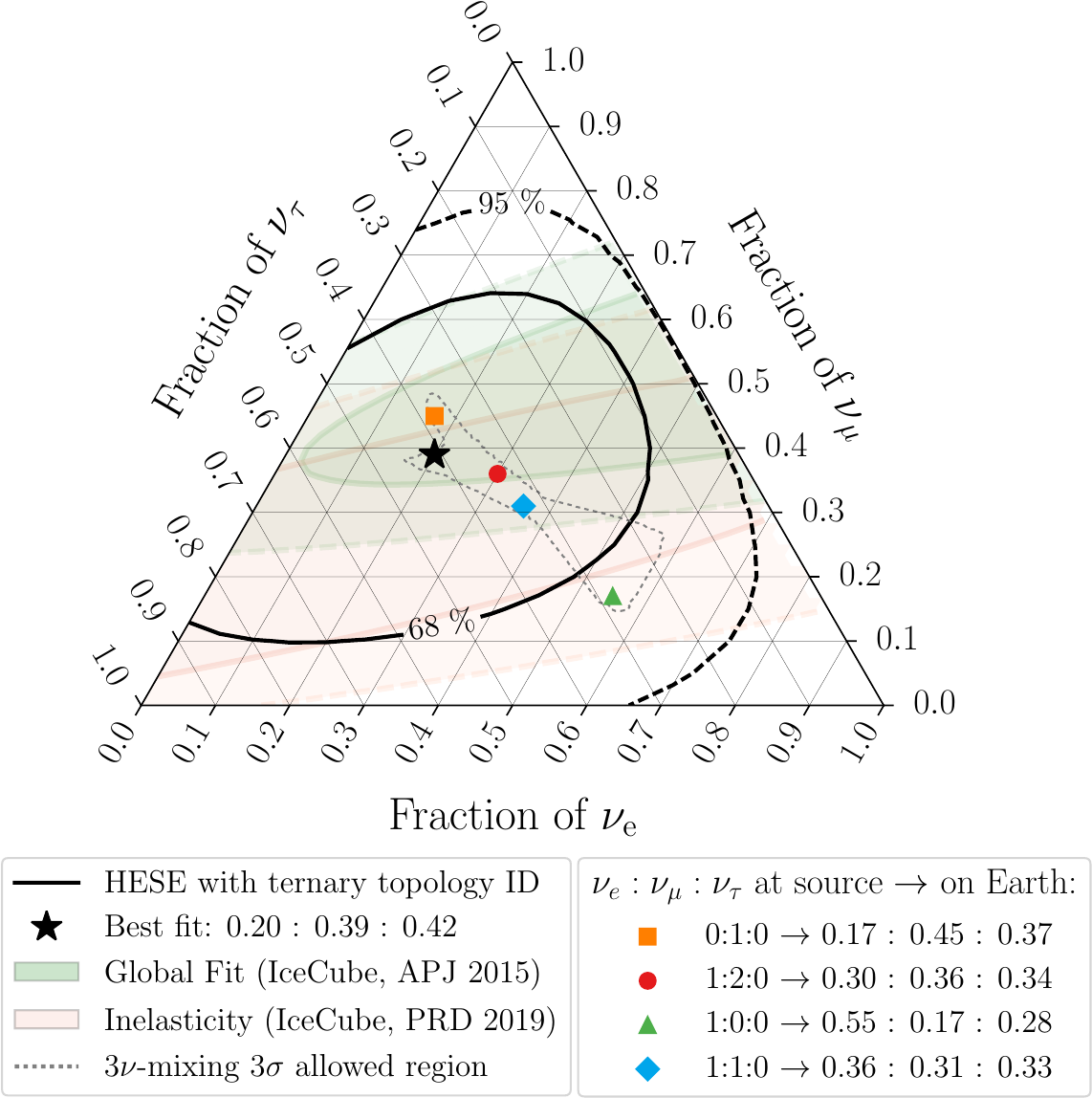}
    \caption{Flavor constraints on the cosmic neutrino flux from various analyses of IceCube data. The constraints derived from the analysis that identified the first tau neutrino candidates~\cite{IceCube:2020fpi} is shown as black contours. Constraints from earlier measurements,  \cite{IceCube:2015gsk,IceCube:2018pgc} are shown as shaded regions. They are compared to different scenarios of neutrino production in astrophysical sources and the full range of possible flavor compositions assuming standard flavor mixing (gray dotted region). Figure adapted from~\cite{IceCube:2020fpi}.}
\label{fig:flavor}
\end{figure}

The IGRB is also a key constraint for multi-messenger astrophysics and provides an important link to neutrino observations. Astrophysical neutrinos, discovered for the first time in the data of the IceCube neutrino telescope in 2013~\cite{IceCube:2013low}, have now been observed for over a decade. 
Most of the neutrinos attributed to an astrophysical origin cannot be related to specific known sources. The distribution of the dominant fraction of these neutrinos is compatible with an isotropic arrival pattern, pointing to an extragalactic origin of the particles. We therefore call it in this paper the isotropic neutrino background (INB) in analogy to the IGRB. \autoref{fig:mm_paradigm} shows the spectrum measured by IceCube using track and cascade datasets, respectively. 
Based on the observed ratio between tracks, cascades and double showers, IceCube can also constrain the flavor composition of the astrophysical neutrino beam, an overview of the constraints derived from different data samples is given in \autoref{fig:flavor}. 
The flavor composition is compatible with production of the neutrinos from charged pions in astrophysical sources, and standard-model flavor oscillations of the neutrinos during their propagation to Earth. A more detailed discussion of the relation of flavor composition measurements and BSM physics can be found in \autoref{sec:bsm_signatures} below.

\autoref{fig:mm_paradigm} also relates the measured intensity of the IGRB to the intensity of the INB and the ultra-high energy cosmic-ray flux. 
Neutrinos and gamma rays are produced simultaneously, and with similar energies, when the pions generated in CR interactions decay. While neutrinos can propagate the universe freely, photons with energies above few tens of GeV will interact with UV to microwave radiation backgrounds in the universe~\cite{Dwek:2012nb}, producing electron-positron pairs that in turn will radiate photons (at lower energies than the initial photon) via inverse Compton interactions in the same intergalactic radiation fields. This is a process commonly denoted as cascading of gamma rays.
The energy of the TeV and PeV photons produced in CR interactions in extragalactic sources will therefore reach Earth in the GeV band, as a contribution to the IGRB observed by Fermi LAT.

The described process has profound implications on the nature of the neutrino sources responsible for the INB. The gamma rays co-produced in these sources would also be responsible for the dominant fraction of the high-energy IGRB, unless the sources themselves are opaque to gamma rays. Unsurprisingly, given that most extragalactic Fermi LAT sources are blazars, also the IGRB above few tens of GeV is dominated by emission from unresolved blazars~\cite{Fermi-LAT:2015otn}. 
Evidence for TeV neutrino emission from blazars has also been found, in particular in relation to the source TXS 0506+056~\cite{IceCube:2018dnn,IceCube:2018cha}. However, the lack of the detection of bright individual blazars with IceCube, limits their cumulative contribution to the total extragalactic neutrino flux to few percent~\cite{IceCube:2016qvd}. This implies that the source populations creating the extragalactic neutrino and gamma-ray skys are largely complementary, and that a significant fraction of the neutrino emission arises from sources (or transients) that are themselves opaque to gamma rays. 

Few such environments exist in nature, early phases of gamma-ray burst (GRB) explosions would be a candidate, but the non-observation of neutrinos correlated to GRBs sets stringent limits (<1\%) on their contribution to the extragalactic neutrino emission~\cite{IceCube:2014jkq}. 
Another class of candidates are accretion disks around supermassive black holes in the center of active galaxies (see, e.g., \cite{Liu:2022cph} for a review). 
Their intense optical to UV radiation, in many cases outshining the radiation of the entire Milky Way, efficiently absorbs gamma rays produced in their vicinity. The first strong evidence for a neutrino point source, the observation of neutrino emission from the nearby active galaxy NGC~1068~\cite{IceCube:2022der} strengthens the case to be made for this type of sources. 
NGC~1068 is almost an order of magnitude brighter in neutrinos than in high-energy gamma rays, pointing to a production of the neutrinos near the black hole event horizon %
and the absorption of the co-produced gamma rays~(e.g.,~\cite{Murase:2022dog}). 
In addition, a correlation analysis between neutrino emission and a large sample of active galaxies finds that the IceCube data is compatible with a predominant production of extragalactic neutrinos in this population~\cite{IceCube:2021pgw}. 
Interestingly, a related transient phenomenon, the tidal disruption of stars in the gravitational field of otherwise quiet supermassive black holes that lead to the formation of temporary accretion disks, have also been linked to the emission of high-energy neutrinos~\cite{Stein:2020xhk,Reusch:2021ztx}.  

Signatures of BSM physics can be imprinted in astrophysical signals in many ways. In the brightness, spectral shape and spatial distribution of sources and diffuse emission, as well as in the temporal characteristics of the emission from transients. 
For neutrinos, the flavor composition is another unique probe. So far all searches have only produced constraints on such new physics processes. In the section below, we will introduce such signatures, before 
discussing the constraints derived in \autoref{sec:new_physics}.

\subsection{Signatures of BSM}
\label{sec:bsm_signatures}
\subsubsection{Sources and transients} %
Our current understanding of physics puts constraints on the acceleration of cosmic rays in astrophysical environments, secondary particle production in astrophysical sources, and the propagation of high-energy particles through the universe.
This can be used to search for new physics by testing the data for violations of such constraints, and -- if none are observed -- calculating limits on key quantities defining the BSM physics considered. 
The intrinsic gamma-ray spectra of extragalactic sources are modified by the absorption of gamma rays in the extragalactic background light (EBL, e.g.,~\cite{Franceschini:2017iwq}) due to photon-photon pair-production of the gamma rays with the microwave-to-UV photons of the EBL. 
A strong, distance dependent, absorption of photons with E$\,\gtrsim\,$10~GeV is expected from this interaction. 
Robust lower limits on the energy density of the background light in the universe can be calculated, by summing up the radiation from stars / dust clouds in galaxies, and the cosmic microwave background (CMB). 
A significant excess of high-energy photons from distant sources could therefore be interpreted as a signature of new physics, in particular of axion-like particles (ALPs) that couple to photons in the presence of external magnetic fields (see, e.g.,~\cite{Raffelt:2006cw}) .

Astrophysical transients give rise to another class of tests for the standard model, i.e., time-of-flight tests for high-energy particles of different energies. Quantum gravity theories generally allow for a violation of the energy dispersion relation of special relativity that would lead to small, energy dependent, differences in the group velocities of particles that deviate from $E^2 - p^2 = m^2$ (see \cite{Martinez-Huerta:2020cut}, for a review on this topic). Photons or neutrinos emitted at the same time and location in an astrophysical source would not arrive at the same time for observers at Earth.  Consequently, upper limits on the arrival time difference can be calculated for particles arriving within a short transient. The enormous distance between source and observer, often of the order of 10$^9$ light years and more, and the wide energy range available in the observations of high-energy particles translates into strong constraints on Lorentz invariance violations. 

\subsubsection{Diffuse emission spectra and neutrino flavor composition} %

We discussed in \autoref{sec:astro_observations} that the annihilation of WIMPs in the dark matter halo of the Milky way might give rise to an observable diffuse gamma-ray signal from the central parts of our Milky Way. However, interpretation of the observed gamma rays is complicated due to a potential astrophysical origin of the emission. 
An easier-to-interpret signal would come from observations of the Milky Way's satellites, called dwarf spheroidal galaxies (dSph).
Due to the lack of gas or dust in such galaxies, and the absence of significant star-formation activities, no astrophysical gamma rays are expected from such galaxies, and an observation would be a ``smoking gun'' for WIMP annihilation. %

The observed spectral shapes of the IGRB and the INB that contain the respective cumulative intensity from all unresolved sources and diffuse emission processes in the universe are another sensitive probe for new physics. Indeed, if WIMPs decay, IGRB and INB provide sensitive probes on their lifetimes that are independent of assumptions about the shape of the Galactic halo. The IGRB and INB can be used to derive limits for WIMP masses in the GeV to PeV energy range (e.g., ~\cite{Ishiwata:2019aet}). Constraints can not only be derived for WIMP decays, but also for other dark matter candidates that eventually produce gamma rays, such as the evaporation of primordial black holes (see, e.g.,~\cite{Villanueva-Domingo:2021spv}, for a review). 

A particularly important probe for new physics is the flavor composition of the INB. Extragalactic neutrinos propagate millions to billions of light years before being detected on Earth, the distance being much larger than the coherence length of the associated wave packets. Therefore, instead of an L/E dependent flavor composition an average flavor content is observed that depends only on the flavor composition that is produced in the neutrino sources. Standard oscillations, parametrized in the PMNS matrix~\cite{Pontecorvo:1957qd,Maki:1962mu}, lead to a narrow range of observable flavor compositions for any possible flavor composition at the source. A deviation from this ``allowed'' range could be caused by a variety of new physics processes, including neutrino decay, mixing with sterile neutrinos, neutrino-dark matter interactions, Lorentz invariance violation, etc. The potential of current and future neutrino telescopes to constrain such processes is briefly discussed below in \autoref{sec:new_physics}, \autoref{sec:neutrino_properties}.

\label{sec:flavorcomp}
\subsubsection{Atmopsheric neutrinos} %
Atmospheric neutrinos produced by cosmic-ray interactions provide a beam for the study of neutrino properties. Neutrino oscillation was discovered through studies of atmospheric neutrinos. They are produced as decay products in hadronic showers resulting from collisions of cosmic rays with nuclei in the atmosphere. 

IceCube has collected a large sample of atmospheric neutrino events in the energy range spanning 100 GeV to 100 TeV which can be exploited for BSM neutrino oscillation studies. The BSM topics addressed with this data include searches for sterile neutrinos, anomalous decoherence e.g. through quantum gravity, Lorentz violation, non-standard interactions and neutrino decay.

\subsubsection{Signatures of new particles} %
If the standard model of particle physics holds up to much higher energies, new particles cannot be created at accelerators but may be observed as heavy relics of the big bang era or from ultra-high energy astrophysical sources. 
Hence, IceCube has developed a rigorous program to exploit the unique capabilities of the detector to search for a variety of Big Bang relics and heavy particles predicted to be produced in high energy processes, e.g., in grand unified theory (GUT) symmetry breaking. 
As an example, most theories describing a unification of forces predict the existence of elementary magnetic charges (Magnetic Monopoles, MM) and hence explain the observed quantization of electric charge. 

BSM physics may also manifest itself at lower energies through couplings not accessible at accelerators. 
Minimal dark matter models provide both a dark matter candidate and can explain the neutrino mass via coupling to the Higgs field (see e.g.~\cite{deBoer:2021pon}).

\subsubsection{Neutrinos from beta decays} %
The squared neutrino mass, $m_{\nu}^{2}$, is obtained from a fit of the calculated analytical spectrum to the measured data. 
The analytical spectrum is a convolution of the theoretical differential spectrum with the experimental response function. The response function describes the number of $\upbeta$ electrons that are transported through the flux tube and counted at the detector, including both the energy loss experienced as the $\upbeta$ electrons propagate through the source and the transmission properties of the MAC-E filter.

Testing physics beyond the standard light-neutrino picture adds a variety of possible scenarios to KATRIN. 
The motivation for many of these scenarios relates to the fact that the neutrino mass is produced in a different way from the other masses of Standard Model (SM) particles, and that neutrinos in general are perfect candidates to be connected to BSM physics. 
Common features of such frameworks are additional sterile neutrino states, new neutrino interactions, new particles coupled to neutrinos, etc. 
The scale at which new physics appears is not clear, especially for sterile neutrinos. 
KATRIN is sensitive to two interesting mass scales, the \si{\electronvolt}-\ scale, which corresponds to the design analysis interval of the experiment, and the \si{\kilo\electronvolt} scale, which falls within the $Q$ value of tritium decay. 
The new physics can manifest itself in the measured spectrum in several ways, e.g.\,a kink-like spectral feature (heavy sterile neutrinos~\cite{Mertens:2014nha}), a shape distortion (exotic weak interactions, light sterile neutrinos, or new light bosons~\cite{Arcadi:2018xdd}), the appearance of a line feature (relic neutrino capture~\cite{KATRIN:2022kkv}), or a sidereal modulation of the fit parameters (Lorentz invariance violation~\cite{KATRIN:2022qou}). 

\section{New physics constraints}
\label{sec:new_physics}
\subsection{Primordial black holes} %

Black holes might have formed in the early universe over a wide range of masses, ranging from the Planck mass (10$^{-5}$ g) to hundreds of thousands of solar masses~\cite{Zeldovich:1967lct,Hawking:1971ei,Carr:1974nx}. These primordial black holes (PBHs) eventually disintegrate via Hawking radiation on time scales that depend strongly on their initial mass~$M$. Their lifetime is about $\tau \sim (M/M_{\solar})^{3} \times 10^{64}$~yr. PBHs with a mass of $M_{*} \sim 10^{15}\,$g would therefore have a lifetime comparable to the age of the universe. While disintegrating, they radiate as black bodies with a temperature of $T \sim (M_{\solar}/M) \times 10^{-7}$~K. This implies a strong rise of their temperature briefly before evaporation, leading to bursts of high-energy radiation from individual black holes, and a contribution to high-energy radiation backgrounds throughout the universe.

For initial masses $M \lesssim M_{*}$, the PBHs already have evaporated in the present-day universe. Their early-universe density can be constrained by comparing the calculated radiation contribution from their evaporation to measurements of the IGRB. PBHs with masses $M \gtrsim M_{*}$ can still exist at the current epoch, and their thermal radiation could be observed both, via their contribution to diffuse radiation backgrounds (IGRB, Galactic diffuse emission), and via searches for individual radiation bursts from individual nearby evaporations of PBHs. However, for masses $M \gg M_{*}$, PBH temperatures at the current age of the universe are too low to emit observable amounts of high-energy radiation, limiting any constraints to a mass range within a few orders of magnitude around $M_{*}$.

The observation of the IGRB leads to one of the best current constraints on the density of PBHs with mass ranges between $M\sim10^{14}\,$g and $M\sim10^{17}\,$g~\cite{Carr:2009jm}. Limits from the non-observation of gamma-ray flares due to individual primordial black holes evaporating near the solar system are currently several orders of magnitude weaker (see, e.g.,~\cite{Fermi-LAT:2018pfs}).

Neutrinos are expected to be produced from Hawking radiation as well as the temperature rises in the final stages of the evaporation. 
This expected flux of high-energy ($> 1$ TeV) neutrinos from an evaporating PBH can be used to place constraints on their burst rate density in our local universe. The expected neutrino signal is time and energy dependent~\cite{Dave:2019epr}.
A point-source search for neutrino signal from PBH bursts using several years of IceCube data is forthcoming.

\subsection{Magnetic monopoles} %
Due to the absence of magnetic charges, Maxwell's equations of classical electrodynamics appear asymmetric.
P. Dirac was the first to speculate about the existence of magnetic monopoles~\cite{Dirac:1931kp}. He showed that the presence of a magnetic monopole with a minimum charge $q_m$ would explain why the electric charge is always quantized. 
Although, as Dirac showed, magnetic monopoles can be consistently described in quantum theory, they do not appear automatically in that framework. 
As was first found independently by 't Hooft~\cite{tHooft:1974kcl} and Polyakov~\cite{Polyakov:1974ek}, this is different in Grand Unified Theories (GUT) which embed the Standard Model interactions into a larger gauge group. These theories are motivated by the observation that the scale-dependent Standard Model gauge couplings seem to unify at very high energies. Generally, a {\it 't Hooft--Polyakov} monopole can arise from the spontaneous symmetry breaking of the GUT group via the Higgs mechanism. The stability of the monopole is due to the Higgs field configuration which cannot smoothly be transformed to a spatially uniform vacuum configuration. 

While inaccessible to collider experiments, very heavy GUT monopoles could have been produced in the early universe, when the temperature exceeded $T_{\rm cr}\sim \Lambda_{\rm GUT}$. In this case, the expected monopole density would be roughly one per correlated volume, which corresponds to the horizon size in a second-order phase transition~\cite{Kibble:1976sj,Kibble:1980mv}. This gives a naive monopole energy density (relative to the critical density) of
\begin{equation}\label{Omono}
  \Omega_{\rm GUT}h^2 \simeq 10^{20}\Big(\frac{T_{\rm cr}}{10^{16}\,{\rm       GeV}}\Big)^3\Big(\frac{M_{\rm m}}{10^{17}\,{\rm GeV}}\Big)\,.
\end{equation}
This prediction is in clear conflict with the observed spatial flatness of the Universe ($\Omega_{\rm tot} \simeq 1$) and known as the ``monopole problem''.
A solution to this problem is an inflationary universe, an exponential expansion of the scale factor diluting the initial monopole abundance.

Although the greatest interest has been in the supermassive monopoles that are motivated by a large variety of GUTs, the possibility of lighter monopoles lately also received attention (see e.g.~\cite{Ellis:2016glu}). 
Therefore, a large parameter space of possible monopole masses, velocities, and signatures is considered experimentally. 
Here we address direct searches for primordial monopoles with IceCube.

\begin{figure*}[t]
\centering
\includegraphics[height=0.3\linewidth]{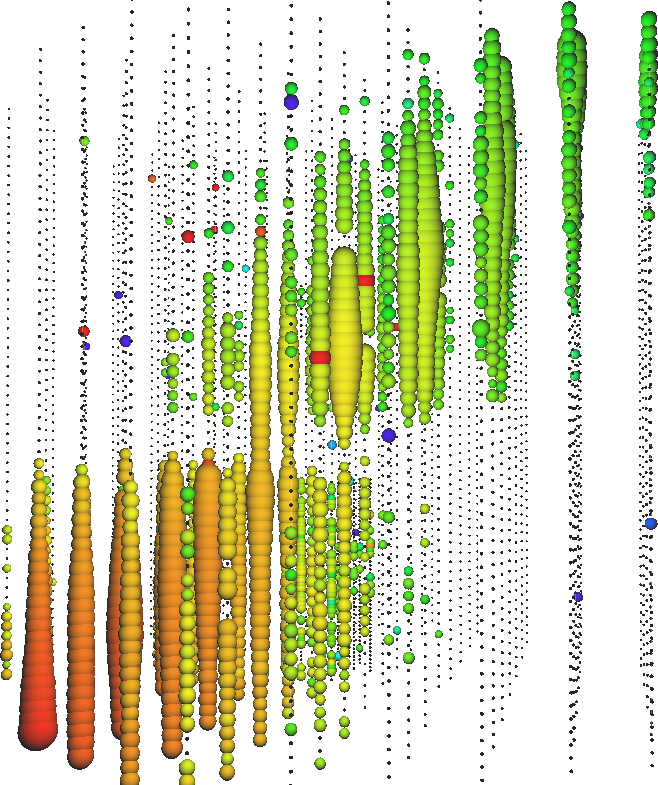}
\hspace{1cm}
\includegraphics[height=0.3\linewidth]{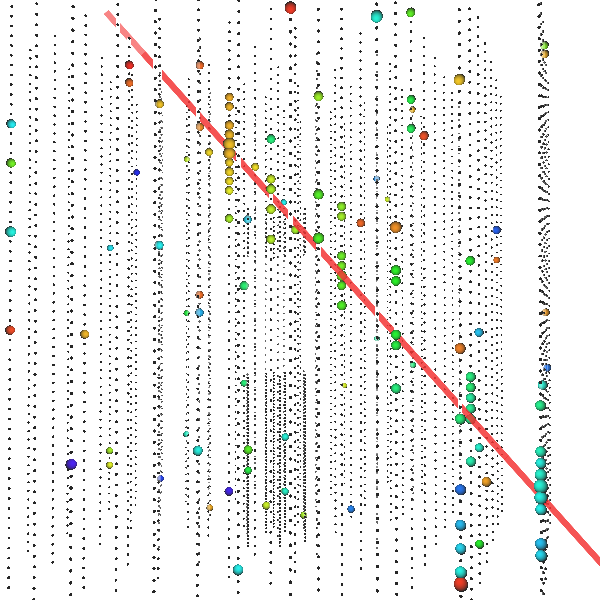}
\hspace{0.5cm}
\includegraphics[height=0.3\linewidth]{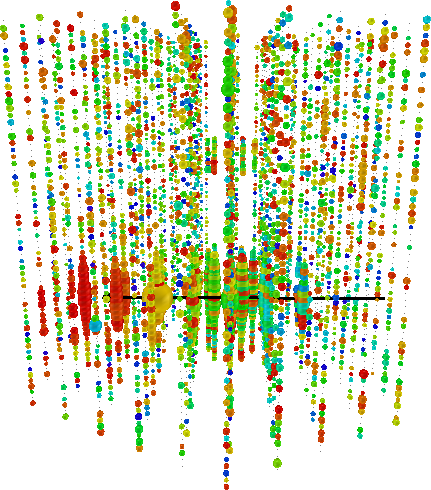}
\caption[]{Simulated event topologies for magnetic monopoles of different velocities. The size of the dot is proportional to the amount of light detected and the color shows the timing: red denotes earlier, blue denotes later hits. Left Panel: monopole traveling at $0.982 c$ from the bottom to the top of the detector~\cite{Lauber:2018ntx}. Middle Panel: monopole at $0.3 c$ moving from the top of the detector to the bottom emitting luminescence light. The simulated track of the particle is indicated in red~\cite{IceCube:2021eye}. Right Panel: Monopole with $\beta = 10^{-3}$ catalyzing nucleon decay with $\lambda_{\mathrm{cat}} = 1\,\mathrm{cm}$ with superimposed background noise~\cite{IceCube:2014xnp}. The black line represents the simulated monopole track.}
\label{fig:MonopolePassage}
\end{figure*}

There are several light-producing channels for the detection of magnetic monopoles in ice, each of which is dominant at different velocities.
Magnetic monopoles can induce Cherenkov light like any other highly electrically charged particle when they travel with velocities above the speed of light in the medium. 
Below the Cherenkov threshold in ice ($\approx 0.76\,\textrm{c}$), no 
direct Cherenkov light is produced.
Instead, indirect Cherenkov light, produced by secondary $\delta$ electrons induced by a passing magnetic monopole, becomes the dominant light production mechanism down to about 0.6\,c.
Unlike for direct Cherenkov light, there is no sharp cutoff. 
Instead, the light yield decreases until luminescence becomes the dominant light production mechanism.
Luminescence light is largely velocity independent, but has a lower overall light yield than the other two light production channels described above.
At velocities well below 0.1 c, luminescence is expected to drop off steeply. Catalysis of nucleon decays is a plausible scenario for GUT monopoles and can be observed when the mean free path
is small compared to the detector size. Examples of signatures of magnetic monopoles passing through the detector at different velocities are shown in Fig.~\ref{fig:MonopolePassage}. 
For each of these speed ranges, the search for magnetic monopoles with the IceCube experiment has set the current best upper limits on the flux of magnetic monopoles~\cite{IceCube:2014xnp,IceCube:2015agw,IceCube:2021wvc,IceCube:2021eye}.

\subsection{Generic searches for Big Bang relics}
A generic experimental approach to search for big bang relics is detecting particles that are much heavier and produce less light than standard model particles. 
Examples for particles producing such signatures are heavy neutral leptons, fractional charges and electrically charged Q-balls which are also candidates for dark matter. 
As the largest and most massive instrumented volume on Earth, IceCube is ideally suited to observe the interaction of such relics. In contrast to seawater, the radioactive purity of South Pole ice allows much lower thresholds for light detection.
With the determination of the efficiency of luminescence in ice~\cite{Pollmann:2021jlo} it is now possible to detect such particles essentially across all velocity ranges.

\subsection{Lorentz invariance violation} %

Quantum gravity theories can lead to a modified dispersion relation and therefore a violation of Lorentz invariance (LIV). The modified dispersion relation can be effectively described as a Taylor expansion:

$$E^2 - m^2c^4 = p^2c^2\,\left(1-\sum_{n=1}^{\infty} s_{\pm} \left(\frac{E}{E_{QG}}\right)^{n} \right)$$.

$E_{QG}$ is the energy scale at which the deviations become prominent, and usually $E_{QG} \sim E_{\mathrm{Planck}}$. This dispersion relation implies an energy-dependent velocity $v$ of the photons (or, e.g., neutrino mass eigenstates) that is in leading order given by $ 1-v/c \propto (E/E_{QG})^n $, where n=1,2,\dots is the lowest order non-vanishing term in the expansion \cite{Amelino-Camelia:1997ieq}.

An obvious cosmic probe for such a modified dispersion is the arrival time of photons of different energies that have been emitted at the same time, or within a known time frame. Due to the long propagation distances involved, even small deviations from the canonical dispersion relation lead to a measureable time lag. However, in most astrophyscial environments the emission time of photons cannot be pinpointed exactly. In the case of violent outbursts, e.g., GRBs, it is reasonable to assume that all photons have been emitted after the onset of the burst and constraints can be derived from the observed time difference. 

In May 2009, the Fermi satellite observed the bright short GRB 090510 with both on-board instruments. The GBM registered the light curve of the burst above 8~keV in energy, while the LAT measured the emission above 100~MeV, including the arrival of a 31~GeV photon about 850~ms after a precursor peak and 300 ms after the onset of the main burst. Together with the large distance of the burst, determined later from optical observations to originate at a redshift of $z \sim 0.9$, this allowed for the first time to constrain $E_{QG} > E_{\mathrm{Planck}}$ for $n=1$~\cite{FermiGBMLAT:2009nfe}.

The most sensitive LIV constraints from neutrinos do not arise from time-of-flight measurements but from  studies of the impact of the modified dispersion relation on flavor oscillations. Muon neutrinos, produced in CR air showers, propagate through the Earth as a superposition of mass eigenstates. The mass differences between the eigenstates lead to the well known phenomenon of flavor oscillation. Modifications in the dispersion relation change the probability $P_{\nu_\mu \rightarrow \nu_\mu}$ for observing a muon neutrino of energy $E$ at distance $L$ from the production location. IceCube measurements of the atmospheric neutrino flux as a function of zenith angle can probe propagation distances $L$ between few tens of km and the Earth diameter of about 12800~km, as well as energies between few hundred GeV and few tens of TeV, leading to some of the strongest constraints for higher-order deviations of the dispersion relation~\cite{IceCube:2017qyp}.

\begin{figure*}[t]
	\centering
	\includegraphics[width=\textwidth]{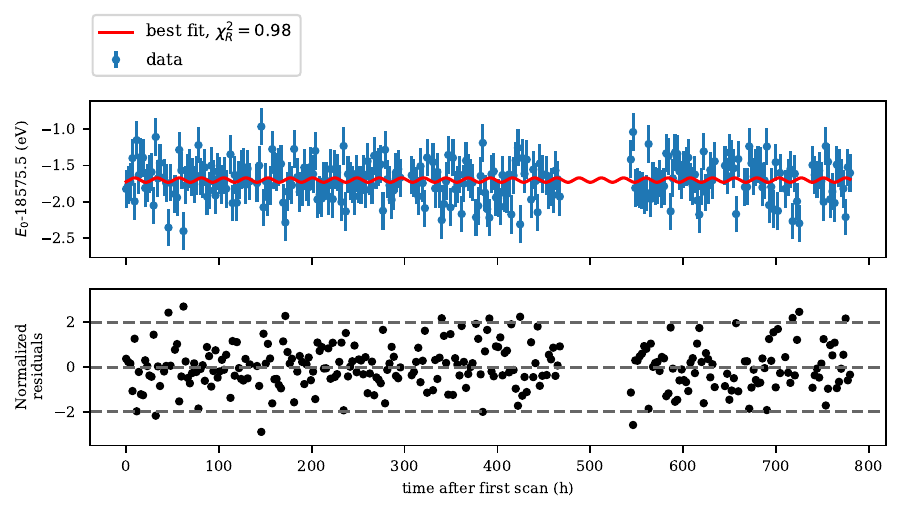}
	\caption{Illustration of the best fit of a sidereal oscillation. The fitted endpoints of the first physics campaign (blue) have been used to fit an oscillation with sidereal frequency and free amplitude and free phase. The bottom panel displays the normalized residuals of the data fit (black). Figure adapted from~\cite{KATRIN:2022qou}}
	\label{fig:bestfit}
\end{figure*}

While many of the possible Lorentz-invariance violating parameters are strongly constrained by time-of-flight or neutrino-oscillation experiments, some parameters are virtually invisible to these experiments (see e.g.~\cite{Diaz:2013saa}). In KATRIN a preferred direction in space could manifest as a variation of the spectral endpoint depending on the latitude and orientation of the $\upbeta$-electron beam, or as a sidereal oscillation of the spectral endpoint $E_0$~\cite{Lehnert:2021tbv}.
KATRIN has performed a search for LIV based on its first science run, looking for an oscillatory signature with the required period of \SI{23.7}{\hour} (figure~\ref{fig:bestfit}).
This analysis finds no sign of LIV~\cite{KATRIN:2022qou}.

\subsection{Dark matter annihilation and decay} %
\label{sec:DM}
The annihilation or decay of dark matter into Standard Model particles can be probed by $\gamma$-ray, cosmic ray, and neutrino emission. Indirect dark matter searches with IceCube using neutrinos cover signals from the Sun, the Earth, galaxies and galaxy clusters. For a recent review see~\cite{Ahlers:2018mkf}.

The Fermi LAT measurements are responsible for some of the strongest constraints on both, the annihilation cross-section of WIMPs and the lifetime of decaying DM in respective models. An excess of GeV radiation is observed from the Galactic center region (e.g.,~\cite{Fermi-LAT:2015sau}) that can be interpreted as a WIMP annihilation signal for a dark matter particle with a mass of few tens of GeV. 
However, an astrophysical origin of this excess due to emission from unresolved gamma-ray source populations is another viable candidate for its origin. 
Interestingly, for dSph galaxies in the vicinity of the Milky way, where astrophysical backgrounds are low, no gamma-ray emission has been found so far, leading to constraints on the annihilation cross-section below the thermal relic cross-section for WIMPs with masses $M_{DM} \lesssim 100\,$GeV, depending on the annihilation channel. 
The work in~\cite{DiMauro:2021qcf} demonstrates that, including also the direct CR measurements from AMS-02~\cite{AMS:2021nhj}, it is indeed difficult to reconcile all current measurements with a dark matter origin of the Galactic center excess.

The Fermi LAT IGRB measurement provides one of the strongest lower limits on the lifetime $\tau$ of decaying dark matter. The limits for several decay channels exceed $\tau \gtrsim (1 - 5) \times 10^{28}\,$ over a DM particle mass range from few GeV to EeV~\cite{Blanco:2018esa}. 
In contrast to the signal from annihilation, which is proportional to the square of the dark matter density, and therefore best visible in overdense regions such as the center of the Milky Way, a signal from decaying dark matter is proportional to the density only. 
In this case the signal integrated over the entire visible universe is competitive, and, in addition, reprocessing of high-energy photons in the radiation fields of the universe to GeV energies, yields a strong signal even for DM particle masses far beyond the energy range of the Fermi LAT.

\subsection{Neutrino properties} 
\label{sec:neutrino_properties}
\subsubsection{Neutrino mass} %
Compared to everyday life, the Planck mass is still small with about $10^{-8}$\,kg. 
However, this corresponds to an energy of $10^{28}$\,eV, an energy which seems to be unattainable at present at terrestrial particle accelerators with $10^{13}$\,eV, and which is also not reached by the highest energies of cosmic radiation of $10^{21}$ eV observed so far. 
At particle energies corresponding to the Planck mass, the De Broglie wavelength $\lambda = h/p$ becomes comparable to the Schwarzschild radius. 
Accordingly, effects of a quantization of gravity would have to be considered.
Up to times of about $10^{-44}$ seconds after the Big Bang, corresponding energy densities prevailed. 
Planck time is therefore the point at which the effects of quantum gravity in other fundamental interactions can no longer be ignored. 
Possibly all fundamental forces are united in this state, but the mechanism of this unification is unknown so far.

Currently, there is no generally accepted theory to describe the Planck era. 
However, the key could lie at the other end of the energy and mass scale, namely in the lightest known elementary particle, the neutrino. 
For decades the neutrino was considered massless however neutrino oscillations prove this wrong.
Neutrinos have a mass less than one millionth of the second lightest particle, the electron. 
The masses of all other known elementary particles are much closer together. 

To generate non-vanishing neutrino masses key properties of the Standard Model must be modified as it is not possible to construct a renormalizable mass term for the neutrinos with the fermionic content and gauge symmetry of the SM. 
Abandoning gauge symmetry or Lorentz invariance appears inappropriate. 
In order to introduce a neutrino mass one must extend the particle content of the model or depart from renormalizability.
The idea of the so-called seesaw mechanism is to ascribe the tiny neutrino masses of less than 1~eV to the existence of a very high mass scale.
The high mass scale originates from the generic hypothesis that new physics (NP) beyond the SM only manifests itself above some high energy scale. 
Between the Planck scale and the vacuum expectation value of the Higgs field ($v = 246$~GeV)
there might exist another fundamental scale $\Lambda_{NP}$. 
The heavier the new scale, the lighter the known neutrinos - hence the name Seesaw mechanism.
With the seesaw mechanism the neutrino mass would be suppressed with $v^2 / \Lambda_{NP}$.
An example of an extension to the SM leading to a seesaw mechanism for neutrino masses are
Grand Unified Theories (GUT, see e.g.~\cite{Gell-Mann:1979vob}), in which all forces except gravity are unified with $\Lambda_{NP} \simeq 10^{16}$~GeV. 
In addition to the known light neutrinos, very heavy right-handed neutrinos are produced.
Beyond this minimal seesaw scenario there exist also variants that connect $\Lambda_{NP}$ to the Planck scale (see e.g.~\cite{Mohapatra:1986aw,Mohapatra:1986bd,Smirnov:2018luj}).

Recently, the KATRIN experiment has set the world-leading direct measurement limit of $m_\nu <$ \SI{0.8}{\electronvolt} (\SI{90}{\percent} C.L.), based on its first calendar year of data taking~\cite{KATRIN:2021uub}. 
The experiment continues to acquire data and is expected to run for several more years. Several promising upgrades, including active background-mitigation measures, are currently in development with the aim of reducing or compensating for the observed KATRIN background, and thereby further improving the neutrino-mass sensitivity towards the 0.2~eV sensitivity goal.

\subsubsection{Sterile neutrinos} %
Right-handed neutrinos are a well-motivated SM extension:
In the simplest version of the Seesaw mechanism the Standard Model is extended by assuming two or more right-handed neutrinos in addition to the known left-handed neutrinos.
Right-handed neutrinos, unlike the known left-handed neutrinos, would not participate in the weak interaction and are therefore called \emph{sterile}. %
Here we discuss sterile-neutrino searches which consider a minimal extension of the standard model by one additional sterile neutrino ($3\nu+1$), associated with a mass eigenvalue $m_4$.

By measuring and characterizing the flux of atmospheric neutrinos in the GeV to PeV energy range, the IceCube neutrino observatory is uniquely positioned to search for such oscillations.
For eV-scale sterile neutrino states a matter-enhanced resonance would result in the near complete disappearance of TeV-scale muon antineutrinos passing through the Earth’s core.
An analysis in reconstructed energy-zenith space assuming a sterile neutrino state with a mass-squared difference between 0.01\,eV$^2$ and 100\,eV$^2$ has been performed.
The result shows consistency with the no sterile neutrino hypothesis~\cite{IceCube:2020phf}.

The signature of an eV-scale sterile neutrino in KATRIN is a kink-like distortion of the $\upbeta$ spectrum, most prominent at $E \approx E_0 - m_4$. 
The sterile-neutrino analyses of the first two measurement campaigns with a sensitivity to $m_4^2\lesssim\SI{1600}{\electronvolt\squared}$ and $\lvert U_{e4} \rvert^2 \gtrsim$ \num{6e-3} did not identify a signal~\cite{KATRIN:2022ith}.
Sterile neutrinos with a mass in the kilo-electronvolt (\si{\kilo\electronvolt}) regime are promising dark-matter candidates.
As for eV-scale sterile neutrinos, the presence of a sterile neutrino in the ~\si{\kilo\electronvolt} mass range would result in a characteristic kink-like signature with a magnitude governed by the mixing amplitude.
With an endpoint of $E_0 =$\SI{18.6}{\kilo\electronvolt}, tritium $\upbeta$-decay permits a search for sterile neutrinos on a mass range of multiple \si{\kilo\electronvolt}. 
As a proof of principle, KATRIN data taken in the commissioning run in 2018 has been used to search for \si{\kilo\electronvolt}-scale sterile neutrinos in a mass range of up to \SI{1.6}{\kilo\electronvolt}~\cite{KATRIN:2022spi}.

\subsubsection{Flavor composition constraints on new physics} %

The flavor composition of a neutrino beam emitted from an astrophysical source depends on the astrophysical environment. Neutrinos produced in the interactions of protons and nuclei with ambient matter emerge predominantly from the decay of charged pions that are followed by muon decays. If both, the muon and the pion, decay before losing a significant amount of energy, e.g., due to magnetic fields present in the source region, this will lead to the canonical flavor composition of $\nu_{e}:\nu_{\mu}:\nu_{\tau} = 1:2:0$ at the source. On their long journey to Earth, the produced neutrinos are subject to flavor oscillations. Since the propagation distance of astrophysical neutrinos is much larger than the coherence length of the wave packets, the flavor composition on Earth, assuming standard neutrino oscillation, depends only on the flavor compositions at the source and the elements of the mixing matrix, but not on propagation distance or energy.   

\begin{figure}
    \centering
    \includegraphics[width=0.32\linewidth]{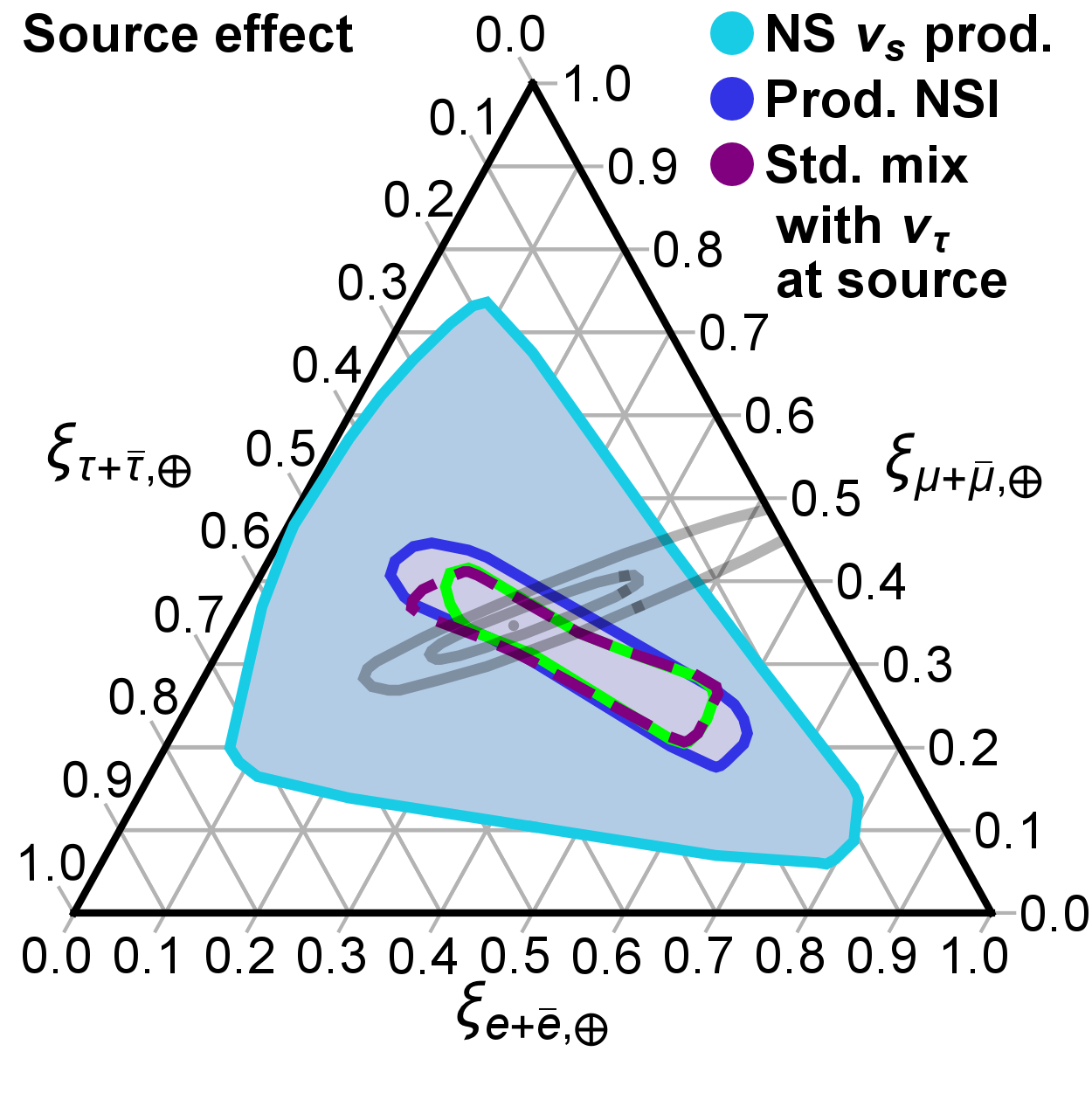}
    \includegraphics[width=0.32\linewidth]{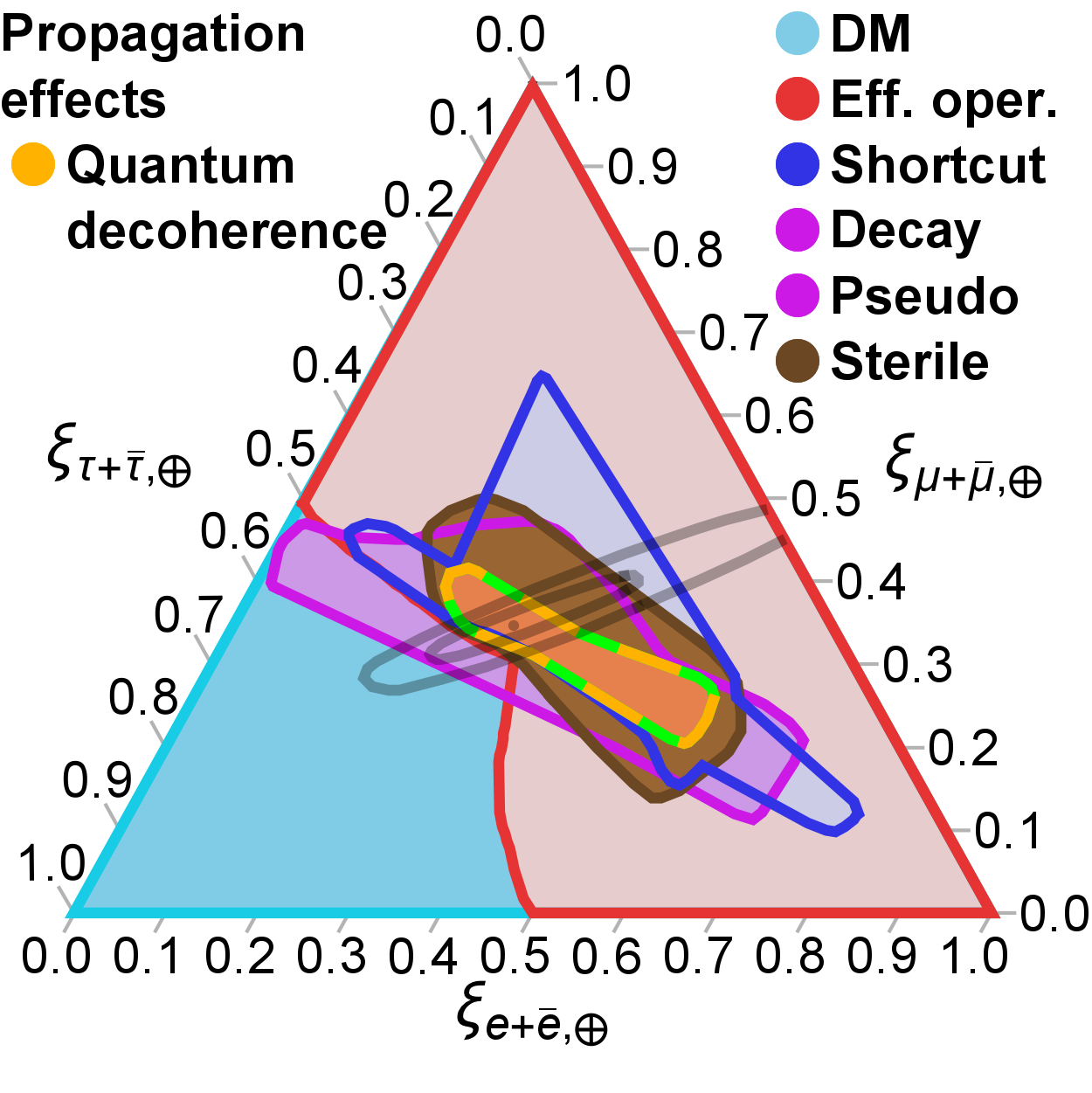}
    \includegraphics[width=0.32\linewidth]{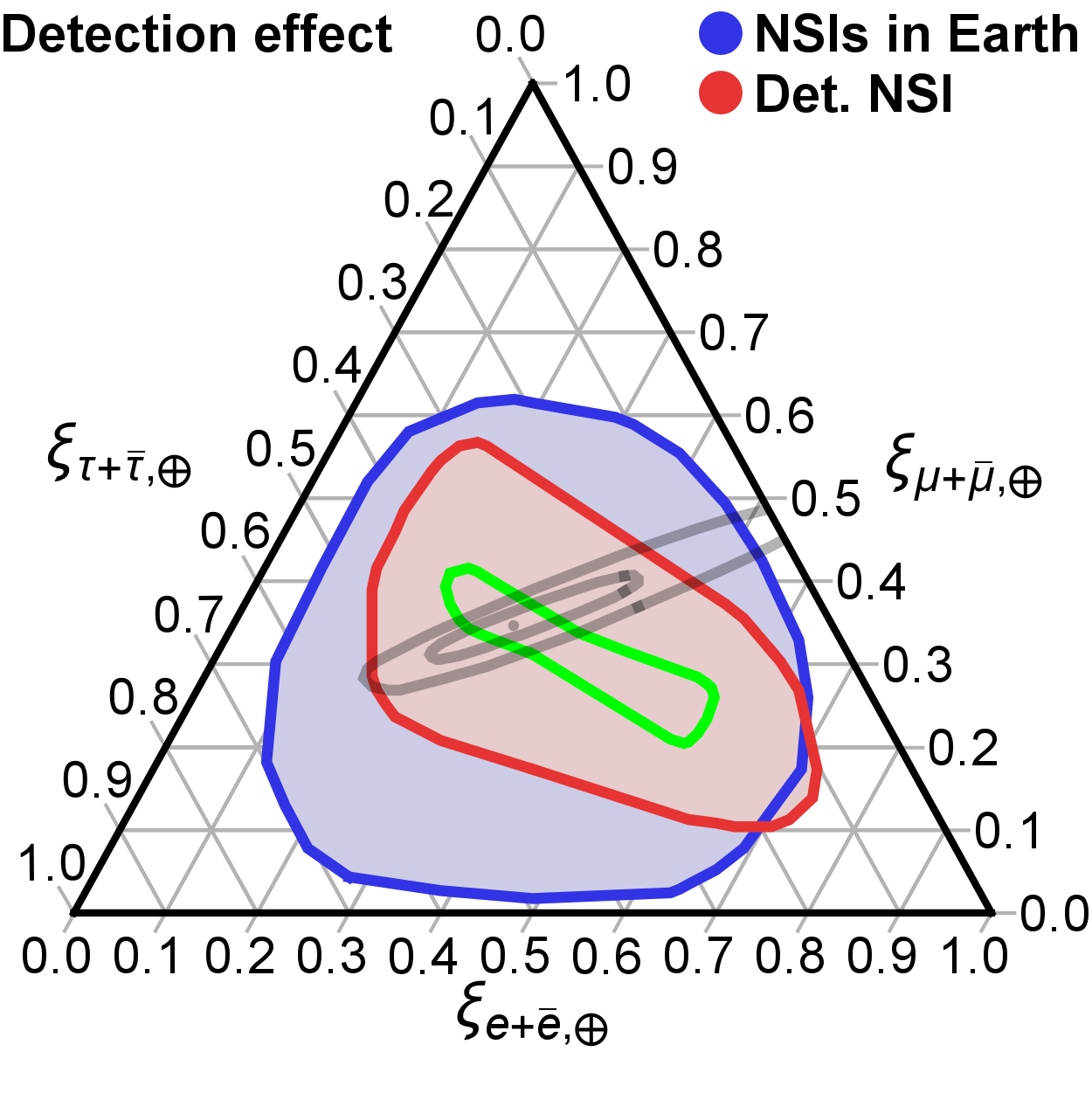}
    \caption{Flavor composition range for selected BSM effects at the source (left panel), during propagation (middle panel), and at detection (right panel). Green regions mark the standard neutrino oscillations expectation, gray contours the IceCube-Gen2 expected sensitivity (best-fit, 1$\sigma$, 3$\sigma$). For more information on the displayed BSM effects, please refer to~\cite{Rasmussen:2017ert}. Figure adapted from~\cite{Rasmussen:2017ert}.}
    \label{fig:flavor_composition_new_physics}
\end{figure}

Indeed, the possible range of observable flavor compositions under standard oscillation assumptions is quite constrained, independent of the flavor ratio at source. A comprehensive overview of the expected flavor compositions with and without new physics effects is given in~\cite{Rasmussen:2017ert}.
\autoref{fig:flavor_composition_new_physics} shows the potential effects of various BSM physics processes, interactions and particles on the astrophysical flavor composition. Some effects manifest from the propagation inside the astrophysical source, some during the propagation of the neutrinos to Earth, and some at the detection stage, when the neutrinos traverse the matter of the Earth. 
The figure also indicates the constraints on the flavor composition that are expected from future detectors, namely IceCube-Gen2 (see \autoref{sec:future}), for the case that the true flavor composition is given by a neutrino production at the source with flavor ratios of $\nu_{e}:\nu_{\mu}:\nu_{\tau} = 1:2:0$ and assuming flavor oscillations according to the best-fit neutrino mixing matrix from atmospheric and accelerator neutrino measurements~\cite{Esteban:2016qun}. Current constraints from IceCube are shown in \autoref{fig:flavor}. 

\section{Future opportunities} %
\label{sec:future}

Since the launch of Fermi LAT in 2008, several ground-based gamma-ray observatories have become operational, or are under construction. The \emph{High-Altitude Water Cherenkov} (HAWC) \emph{Observatory} and the \emph{Large High Altitude Air Shower Observatory} (LHASSO) are both operational and target TeV gamma rays. Both have already delivered a wide variety of new physics constraints (e.g.,~\cite{HAWC:2019wla,HAWC:2019gui,LHAASO:2021opi,LHAASO:2022yxw}) and will continue to do so in the future. The under construction \emph{Cherenkov Telescope Array} (CTA), will allow observations of gamma-ray sources at energies between few tens of GeV and few hundred TeV with unprecedented sensitivity, promising, e.g., strong constraints on the WIMP annihilation cross-section beyond the mass range that can be tested by Fermi LAT and even more sensitive LIV tests~\cite{CTAConsortium:2017dvg}. 

KATRIN has a design sensitivity goal of \SI{0.2}{\electronvolt} at \SI{90}{\percent} confidence. 
This goal was based on a set of projections about achievable backgrounds and systematic uncertainties.
Several systematics 
are exceptionally well controlled %
while others, not originally anticipated, are now particularly important. 
KATRIN's countermeasures against the dominant backgrounds of prior generation experiments have been largely successful -- but previously unforeseen backgrounds far exceed the original design budget. 
KATRIN is investigating active and passive transverse filters, to distinguish background Rydberg electrons from signal $\upbeta$s. \si{\tera\hertz} radiation is a possible means of actively stimulating the de-excitation of Rydberg atoms, reducing the number of background electrons produced. 
Hence, \si{\tera\hertz} radiation is used to manipulate the atomic states of the Rydberg atoms to reduce the ionization process by thermal radiation.

The Project 8 collaboration~\cite{Project8:2022wqh} performs cyclotron radiation spectroscopy of an atomic tritium source with the goal to reach a neutrino mass sensitivity of 40 meV.
HOLMES~\cite{HOLMES:2019ykt} intends to perform a calorimetric measurement of the energy released in the decay of \textsuperscript{163}Ho with a with a sensitivity on the neutrino mass of about $1$\,eV/c$^2$. ECHo~\cite{Gastaldo:2017edk} uses arrays of Metallic Magnetic Calorimeters with \textsuperscript{163}Ho implanted and aims to reach sub-eV sensitivity.

KM3NeT~\cite{KM3Net:2016zxf} is a neutrino telescope under construction in the Mediterranean sea at two sites. Off the coast of Sicily, the ARCA array is optimized for astrophysics with TeV neutrinos and will eventually reach about 1 km$^3$ in instrumented volume. The second array, ORCA, in the deep sea near Toulon, France, is focused on lower-energy neutrinos and the study of fundamental neutrino properties, such as the mass hierarchy or oscillation parameters. Both detectors will complement and improve the current IceCube constraints on BSM physics processes.

\begin{figure}
    \centering
    \includegraphics[width=0.99\linewidth]{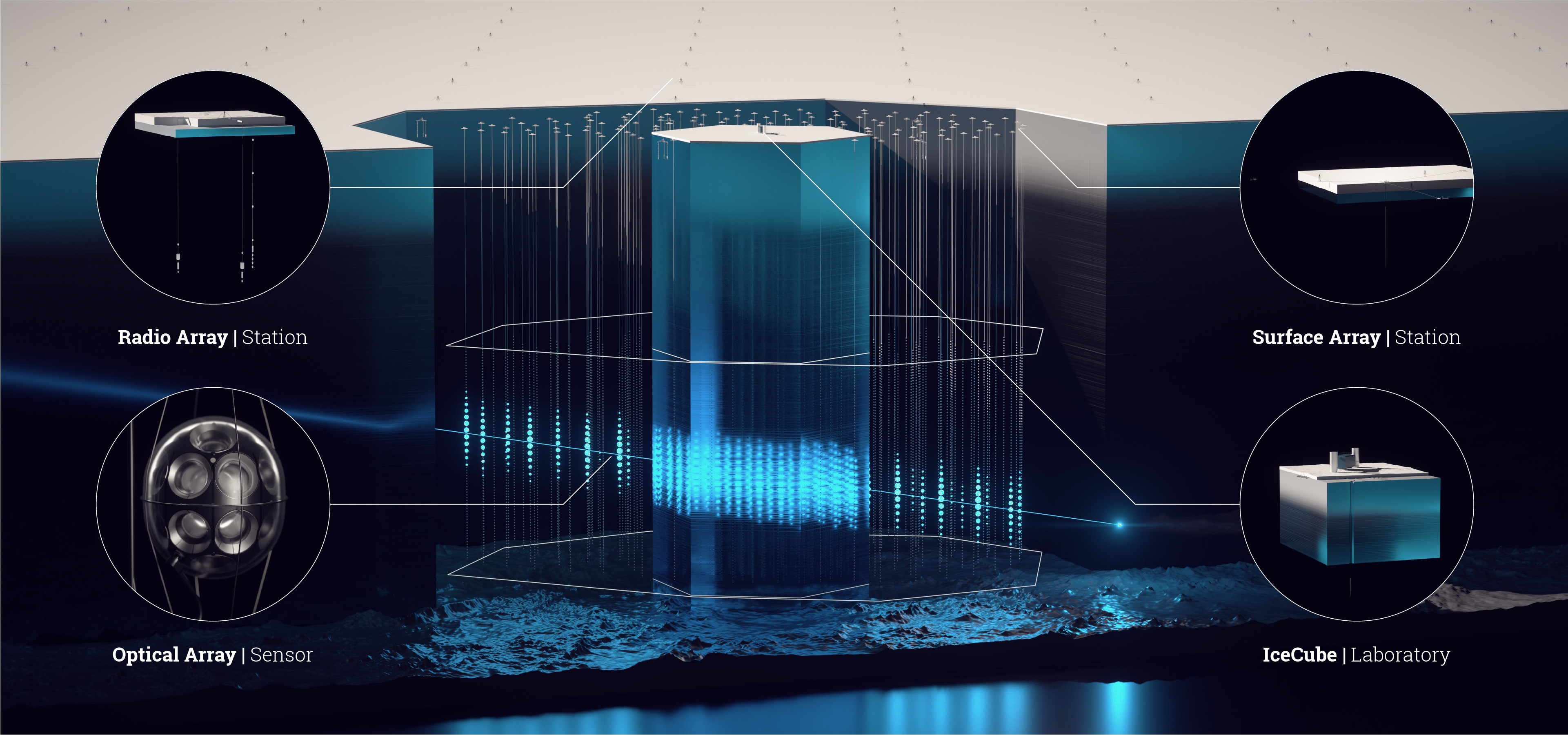}
    \caption{Artist impression of the future IceCube-Gen2 detector. The current IceCube detector will be augmented and extended by seven new strings with optical sensors in the center of IceCube (IceCube Upgrade); 120 new strings with optical sensors extending the instrumented volume to 8 km$^3$ (Gen2 optical array); a surface CR detector covering the footprint of the optical array (Gen2 surface array) and an $\sim 500\,$km$^2$ array (Gen2 radio array) of antenna stations for the radio-detection of ultra-high-energy neutrinos ($> 100\,$PeV). The optical sensors designed for IceCube Upgrade, and the Gen2 optical array will feature multiple PMTs per sensors and collect 3-4 times more light than an IceCube sensor~\cite{IceCube-Gen2:2020qha}.}
    \label{fig:gen2}
\end{figure}

\autoref{fig:gen2} shows an artist impression of the next generation detector at South Pole, IceCube-Gen2, that has been proposed~\cite{IceCube-Gen2:2020qha}. As a first step towards this future detector the IceCube Upgrade is currently underway, aiming to deploy seven new strings with newly developed optical sensors and additional calibration devices in the center of IceCube~\cite{Ishihara:2019aao} by 2026. Further steps will increase the instrumented volume of the optical array by a factor of eight, leading to a sensitivity improvement that allows IceCube-Gen2 to detect five times fainter sources than IceCube, and collect almost an order of magnitude more events for spectral and flavor composition analysis. A glimpse at the expected performance of this next-generation detector for flavor composition studies was already shown in \autoref{fig:flavor_composition_new_physics}. 
Likewise, sensitivity to massive Big Bang relic particles and dark matter will be enhanced by this increase in monitored volume.
A more comprehensive overview can be found in~\cite{IceCube-Gen2:2020qha}. Finally, the radio array of IceCube-Gen2, will push the sensitivity for EeV neutrinos by more than an order of magnitude in comparison to current detectors. An observation of such high-energy neutrinos would allow to probe, e.g., neutrino nucleon cross-sections in a completely new energy regime. Instead of optical sensors the radio array uses antennas at a depth of 150~m, as well as close to the surface that can detect the radio emission from a $> 100\,$PeV neutrino interaction over large distances ($> 1~km$). This enables instrumenting a vast volume of glacial ice on a scale that would have been impossible with optical sensors. 

\section*{Acknowledgment}
The authors are grateful for the support of the Alexander von Humboldt Foundation throughout their careers and in particular the support of the Humboldt Kolleg "Clues to a mysterious Universe" which enabled this cooperation.

\bibliography{literature}
\bibliographystyle{plain}
\end{document}